\documentclass[twoside]{article}
\usepackage{amstex}
\usepackage{fleqn}
\usepackage{espcrc2}
\usepackage{epsf}

\newcommand\Sphi{\sin\frac{\phi}{2}}
\newcommand\half{\frac{1}{2}}
\newcommand\Tr{\operatorname{Tr}}
\newcommand\np[3]{{Nucl.\ Phys.}\ #1 (#2) #3}
\newcommand\pl[3]{{Phys.\ Lett.}\ #1 (#2) #3}
\newcommand\prev[3]{{Phys.\ Rev.}\ #1 (#2) #3}

\title{Putting centre dominance under the microscope}

\author{P.W. Stephenson\address{Dipartimento di Fisica and INFN,
             Universit\`a degli studi di Pisa}}

\begin{document}
\begin{abstract}
  We make various short points on the phenomenon of centre dominance
  in SU(2).  The Z(2) dominance seen in Wilson loops is related to the
  loop distribution and to half-odd-integer representations of the
  group.  The distributions also make it clear that, in this picture,
  the requirement of vortices for confinement is essentially trivial.
  We confirm that the same effect appears in the positive
  plaquette model.  The simple random vortex picture is shown to give a
  substantial fraction of the string tension.
\end{abstract}

\maketitle

\section{INTRODUCTION}

There is now considerable evidence that vortices of Z($N$) flux are a
useful way of looking at confinement in $SU(N)$ gauge theories.  The
centre Z($N$) naturally has a special role in the theory; the
corresponding vortices can easily be shown to cause confinement in a
simple model with minimal mathematical baggage.

The nature of real, physical vortices is less trivial, and their
relationship to the other mechanisms proposed for confinement still
only partly understood.  The formation of vortices is due to the
tradeoff between action and entropy; as argued by 't Hooft some time
ago~\cite{Ho78}, it is a dynamical question about the phase structure
of the theory whether they are realised under the prevailing physical
conditions.  If the phase is the appropriate one, entropic arguments
suggest that a vortex should form on some scale around that of the
correlation length, while if it has a finite extension the
contribution to the action can be small.  The path-ordered product of
fields around the outside gives $\exp(2\pi i/N)$.

An important consequence of this is that vortices are related to the
homotopy of the quotient group SU($N$)/Z($N$); it is too simplistic to
consider the Z($N$) part alone as being responsible for the behaviour.
This is enough to invalidate the old claim that centre vortices cannot
explain the string tension seen in the adjoint representation of
SU(2), which has just the symmetry SU(2)/Z(2).  A specific mechanism
for understanding the behaviour in higher representations has been
proposed~\cite{FaGr97}.

Much of the evidence that vortices thought of in this way are genuine
physical objects has come from `projection vortices' in
SU(2)~\cite{DDFa97a}, where the field is reduced to its Z(2)
components.  Here one finds the effect of confinement remains (though
of course this claim is made of various other pictures).  The vortices
have excellent scaling properties, and present some hope for
understanding the behaviour of SU(2) representations higher than
$j=1/2$.

It has been suggested~\cite{KoTo97} that the sign of the Wilson loop
in SU(2), or its projection onto Z($N$) for general SU($N$), is a
counter for vortices: in the SU(2) case an even (odd) number of
vortices pierce the loop if the trace of the loop is positive
(negative).  It was shown that this indeed reproduces the heavy quark
potential --- so well, in fact, that the dynamics here clearly
contains not just confinement, but everything else too, so that
despite the simplicity of the picture it is harder to investigate the
physics of confinement directly; in other words, `vortices' is
intended here in a broader sense than just those objects causing a
linear potential between quarks.  However, it is another hopeful sign
that Z($N$) effects are the important ones, and therefore that the
more specific way of looking at vortices described above will
encapsulate the physics.  Here, we shall investigate the causes of
this result for the Wilson loop.  This is a summary of results
presented in ref.~\cite{me}.

\section{DISTRIBUTIONS AND REPRESENTATIONS}

\begin{figure}[thb]
  \vspace*{-0.4cm}
  \hspace*{-0.3cm}
  \epsfxsize=7.7cm \epsfbox{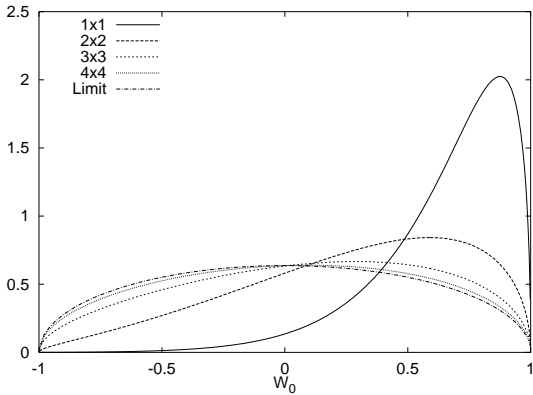}
  \vspace*{-0.9cm}
  \caption{Normalised fundamental Wilson loop distribution at
    $\beta=2.5$.}
  \vspace*{-0.5cm}
  \label{fig:wl_dist}
\end{figure}

One (hitherto unregarded) way of looking at the properties of Wilson
loops is via their distribution.  We define $\rho(W_0(A))$ to be the
normalised distribution of the trace of the Wilson loop $W_0$ of area
$A$ in the fundamental representation such that its integral over
$-1<W_0<1$ is unity. This is shown in figure~\ref{fig:wl_dist} for
various small loops.  For loops much larger than the correlation
length, the value corresponds to a random walk in the gauge manifold,
for which the distribution (shown as the lowest curve) is
\begin{equation}
  \rho(W_0(A\gg \Lambda_\mathrm{QCD}^2)) = \frac{2}{\pi}(1-W_0^2)^{1/2}.
  \label{eq:limiting}
\end{equation}
The distribution allows us to make an important point simply:  if
Wilson loops count vortices, eliminating vortices
trivially removes confinement, as we have only the right hand half of
the distribution where no exponential decay to zero is possible.

\begin{figure}[thb]
  \vspace*{-0.4cm}
  \hspace*{-0.3cm}
  \epsfxsize=7.7cm \epsfbox{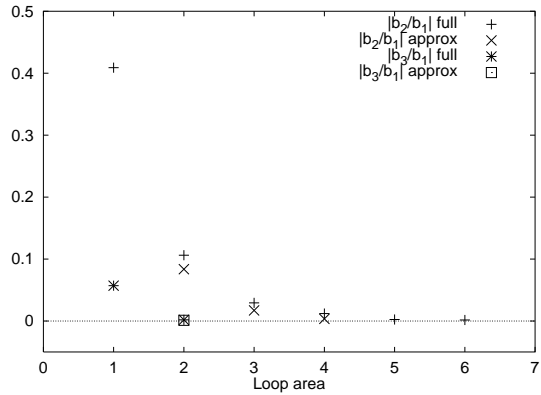}
  \vspace*{-0.9cm}
  \caption{Ratios of Fourier coefficients at
    $\beta=2.5$.}
  \vspace*{-0.5cm}
  \label{fig:fc_rat}
\end{figure}

We shall analyse this distribution via a Fourier analysis,
\begin{equation}
  \rho(\Sphi) = 
  \sum_{n=1}^{\infty} (a_n\cos (n-\half)\phi + b_n\sin n\phi),
  \label{eq:fourier}
\end{equation}
where the units $W_0 \equiv \Sphi$ have been chosen so that
$-\pi<\phi<\pi$.  This form is such that $\rho(\pm\pi) \equiv 0$.  The
expectation value of $W_0$ is easily found to be
\begin{equation}
  \langle W_0\rangle = 3b_0/\pi,
\end{equation}
We can similarly find the expectation value in the centre dominance
picture by assuming a value $-1$ where $\rho<0$ and $1$ where
$\rho>0$. In this case,
\begin{equation}
  \langle W_0^{Z(2)}\rangle = \sum \frac{nb_n}{n^2-1/4}.
  \label{eq:z2expect}
\end{equation}
The formulae differ in two ways: firstly by the presence of all odd
terms for $n>1$ in (\ref{eq:z2expect}), and secondly by an overall
factor of $3\pi/16$ which disappears in ratios and is therefore
unimportant.  The first is investigated in figure~\ref{fig:fc_rat},
where we plot $\lvert b_2/b_1\rvert$ and $\lvert b_3/b_1\rvert$.
Shown in the same graph are the corresponding values from a simple
approximation, where we have selected loops at random from the
plaquette distribution and multiplied them as if uncorrelated: this
corresponds to an area law $\langle \Tr_j(A)\rangle = {\Tr_j({\rm
plaquette})}^A$ for all representations $j$.

\subsection{Representations}
The above clearly demonstrates centre dominance in this sense, but we
can go further.  It turns out that the Fourier series
(\ref{eq:fourier}) corresponds term by term with the representations of
the gauge group, namely, 
\begin{equation}
  \langle\Tr_{n-\half}(W_0)\rangle = \frac{(-1)^{n-1}\pi b_n}{4n}
\end{equation}
for half-odd-integer representations, and
\begin{equation}
\langle\Tr_{n-1}(W_0)\rangle = \frac{(-1)^{n-1}\pi a_n}{4(n-1/2)}
\end{equation}
for the rest.  Note that the $a_0$ term is the trivial identity
representation, corresponding to the distribution
(\ref{eq:limiting}).  So centre dominance results simply from the fact
that the higher representations for $n-\half$, $n>1$ decay more
quickly than the rest.  In particular, if the scaling is Casimir-like,
so that the string tension is proportional to $j(j+1)$ for
representation $j$, the $3/2$ contribution would be expected to decay
exponentially 5 times faster than the $1/2$.  The central issue of the
character expansion was noted in ref.~\cite{AmGr98}.

\section{RANDOM VORTICES}

As has been known for a long time, arguably the simplest way of
generating confinement in SU(2) is to fill the vacuum randomly with
vortices which flip the sign of a Wilson loop.  This can be shown
without reference to the lattice: take an area $A$ embedded in a much
larger area $A'$, and throw vortices into $A'$ up to the required area
density $p_A$.  For non-interacting vortices where perimeter effects
are ignored, the distribution of such objects within the area $A$ is a
simple binomial and gives an area law for the Wilson loop with string
tension $K=2p_A$ as $A'$ is allowed to tend to infinity.  The value of
$p_A$ found for projection vortices (in an indirect projection, and
therefore possibly not optimal) in~\cite{LaRe97} was
$1.9\pm0.2\>\text{fm}^{-2}$ for $K=440\>\text{MeV}$, or around 3/4 of
the required value.  This density is found to have excellent scaling
behaviour~\cite{LaRe97,GrLat98}.

\section{POSITIVE PLAQUETTE MODEL}

\begin{figure}[thb]
  \vspace*{-0.4cm}
  \hspace*{-0.3cm}
  \epsfxsize=7.7cm \epsfbox{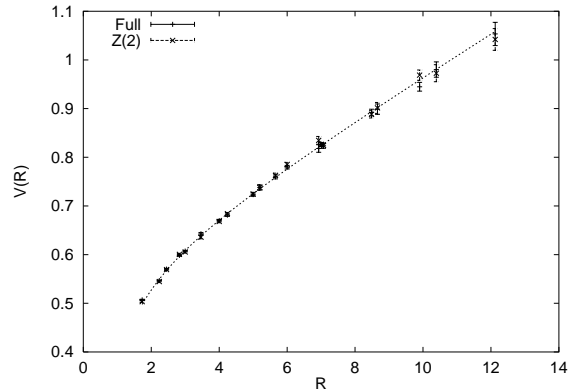}
  \vspace*{-0.9cm}
  \caption{Heavy quark potential in the positive plaquette model at
    $\beta=1.9$.}
  \vspace*{-0.5cm}
  \label{fig:ppm1.9}
\end{figure}

The positive plaquette model~\cite{MaPi82} is an alternative
regularisation of SU(2) in which negative action is forbidden.  It has
all the physical features of SU(2)~\cite{FiHe94}.  Its coupling is
renormalised so that $\beta=1.9$ is slightly weaker than $\beta=2.5$
in ordinary SU(2).  We can use it to show that the centre dominance
effect in the Wilson loop does not require negative plaquettes.
Figure~\ref{fig:ppm1.9} shows the heavy quark potential in this
picture compared with that where only the sign of the loops is taken.
It is clear that the centre dominance effect is present here too.
Another talk at this conference presented similar results~\cite{StLat98}.

\section{CONCLUSIONS}

More work needs to be done connecting vortices with other pictures of
confinement; some results exist in the case of monopoles~\cite{DDFa97b}.
Also, the case of SU(3) needs to be explored further.

\end{document}